\documentclass[aps,pre,twocolumn,superscriptaddress]{revtex4}

\usepackage{graphicx}

\begin{document}

\setlength\tabcolsep{10pt}.

\title{Effects of pyrolysis temperature on the hydrologically relevant porosity of willow biochar}

\author{Jari Hyv\"aluoma}
\affiliation{
Natural Resources Institute Finland (Luke), Tietotie 4, FI-31600 Jokioinen, Finland
}

\author{Markus Hannula}
\affiliation{
BioMediTech Institute and Faculty of Biomedical Sciences and Engineering, Tampere
University of Technology, Tampere, Finland
}

\author{Kai Arstila}
\affiliation{
University of Jyvaskyla, Department of Physics, P.O. Box 35, FI-40014 University of
Jyvaskyla, Finland
}

\author{Hailong Wang}
\affiliation{
Biochar Engineering Technology Research Center of Guangdong Province, School of Environment and Chemical Engineering, Foshan University, Foshan, Guangdong 528000, China; and\\
Key Laboratory of Soil Contamination Bioremediation of Zhejiang Province, Zhejiang A \& F University, Hangzhou, Zhejiang 311300, China
}

\author{Sampo Kulju}
\affiliation{
Natural Resources Institute Finland (Luke), Latokartanonkaari 9, FI-00970 Helsinki, Finland
}

\author{Kimmo Rasa}
\affiliation{
Natural Resources Institute Finland (Luke), Tietotie 4, FI-31600 Jokioinen, Finland
}

\date{\today}

\begin{abstract}
Biochar pore space consists of porosity of multiple length scales. In direct water holding applications like water storage for plant water uptake, the main interest is in micrometre-range porosity since these pores are able to store water that is easily available for plants. Gas adsorption measurements which are commonly used to characterize the physical pore structure of biochars are not able to quantify this pore-size range. While pyrogenetic porosity (i.e. pores formed during pyrolysis process) tends to increase with elevated process temperature, it is uncertain whether this change affects the pore space capable to store plant available water. In this study, we characterized biochar porosity with x-ray tomography which provides quantitative information on the micrometer-range porosity. We imaged willow dried at 60 $^\circ$C and biochar samples pyrolysed in three different temperatures (peak temperatures 308, 384, 489 $^\circ$C, heating rate 2 $^\circ$C min$^{-1}$).  Samples were carefully prepared and traced through the experiments, which allowed investigation of porosity development in micrometre size range. Pore space was quantified with image analysis of x-ray tomography images and, in addition, nanoscale porosity was examined with helium ion microscopy.  The image analysis results show that initial pore structure of the raw material determines the properties of micrometre-range porosity in the studied temperature range. Thus, considering the pore-size regime relevant to the storage of plant available water, pyrolysis temperature in the studied range does not provide means to optimize the biochar structure. However, these findings do not rule out that process temperature may affect the water retention properties of biochars by modifying the chemical properties of the pore surfaces.

\end{abstract}


\maketitle

\section{Introduction}

Pyrolysis biochars are carbon-rich porous materials produced by heating biomass in the absence of oxygen. Porosity and pore size distribution of biochar depend on the raw material selection and process conditions (e.g., type of pyrolysis device, heating rate, maximum temperature and holding time at maximum temperature). Application of biochars to soils can alter and improve soil physical and chemical properties, such as hydraulic properties including hydraulic conductivity and water retention capacity (e.g. Refs. \cite{ref1,ref2,ref3}). Biochar pore sizes can cover a wide range of scales. For direct water holding effects, the minimum pore diameter of interest is approximately 200 nm, which corresponds to the permanent wilting point with matric potential -1.5 MPa \cite{ref4}. Water stored in pores smaller than this is not plant available. In addition, water stored in submicrometre pores above the permanent wilting point is not readily available for plants. For example, it has been reported that dry matter production of plants may virtually cease far before the permanent wilting point \cite{ref5}. Therefore, the characterization of biochars aimed at soil amendment purposes should be focused on the micrometre-scale pores that provide water storage capacity and fast release of water to plants.

Physical characteristics of biochars are often studied by gas adsorption techniques accompanied by the Brunauer-Emmett-Teller (BET) modelling to determine the specific surface area \cite{ref6} or Barrett-Joyner-Halenda (BJH) modelling to determine the pore size distribution \cite{ref7}. Pore space analysis based on these methods and models is limited to pores smaller than 300 nm, whereby gas adsorption measurements do not tell much about the porosity in the size range that is important for biochars that are used as soil amendments to improve the water holding properties of soil \cite{ref8}. In spite of this limitation, in many cases BET analysis still is the only characterization method used for physical pore space even for biochar produced for soil amendment purposes. Mercury intrusion porosimetry is able to quantify the micrometre-scale porosity of porous materials. Unfortunately the feasibility of the method is reduced by the assumptions used in interpretation of the results, which do not hold for arbitrary porous materials \cite{ref9}.

Numerous studies have concluded that elevated pyrolysis temperature leads to increased BET specific surface area (see, e.g., \cite{ref10,ref11,ref12,ref13,ref14,ref15}). Microstructural evolution of biochars during carbonization was studied by Kercher and Nagle \cite{ref16}, who explained that the increase of specific surface area results from the creation of nanometre-range pyrogenetic pores due to the growth of high-density turbostratic crystallites. The total porosity on the other hand does not have a clear dependence on the production temperature. Gray et al. observed that total porosity remained relatively stable with increasing process temperature, which suggests that pyrogenetic nanoporosity does not have significant volume even though it can provide major part of the surface area \cite{ref13}. 

Biochar structure is summarized in the model proposed by Gray et al. \cite{ref13}. In this model the pore space of biochar consists of stable residual porosity resulting from the cellular structure of the raw material, pyrogenetic nanoporosity, and aliphatic functionality on the walls of the residual porosity. Residual pore structure is considered stable and independent of the pyrolysis temperature and forms majority of the total pore volume. Pyrogenetic nanopores are increasingly created as process temperature increases, but even at higher temperatures they form only a minor part of the total pore volume. Aliphatic functionality is volatilized at higher temperatures which results in reduced hydrophobicity.

It is thus likely that feedstock selection is of major importance when one is interested in water holding applications of biochar such as soil amendments. Cellular structure of the feedstock accounts for major part of pore volume and these pores typically are in the size range capable to store readily plant-available water. Pyrogenetic pores are increasingly created with elevated process temperature, but water in these pores is too tightly bound to be available for plants and the total volume of these pores is low. However, it is not clear if process temperature contributes to water holding also by altering the physical pore characteristics, e.g. due to shrinkage in the pore size \cite{ref17}, or only by changing the surface chemistry and consequently the contact angle at the pore walls \cite{ref13}. 

Characterization of porosity development during pyrolysis is still mostly based on indirect measurements and conceptual models rather than on direct evidence. In this paper we report results of an imaging study where micrometre-scale porosity of biochars is directly imaged with x-ray computed microtomography. The three-dimensional image analysis results obtained for willow biochars pyrolysed in different temperatures are compared against gas adsorption measurements. Further information about biochar porosity is obtained with helium ion microscopy (HIM), which is a novel imaging technique within the family of scanning beam microscopes. HIM results in images with very high resolution and large field of depth, thus giving accurate information of submicrometre porosity. HIM also allows for direct imaging of non-conductive samples as the beam induced charging can be neutralized with an electron flood gun. These imaging techniques allows us to get direct evidence on the significance of process temperature on the porosity that is relevant in applications based on water holding capacity of biochar and also more generally in any application relying on the micrometre-range porosity of biochar (e.g. microbial habitats \cite{ref18}). In addition, we are able to compare the importance of pore size regimes probed by gas adsorption techniques in biochar research targeting to soil amendment applications.

\section{Methods}

\subsection{Sample preparation}

\begin{figure}
\includegraphics[width=0.45\textwidth]{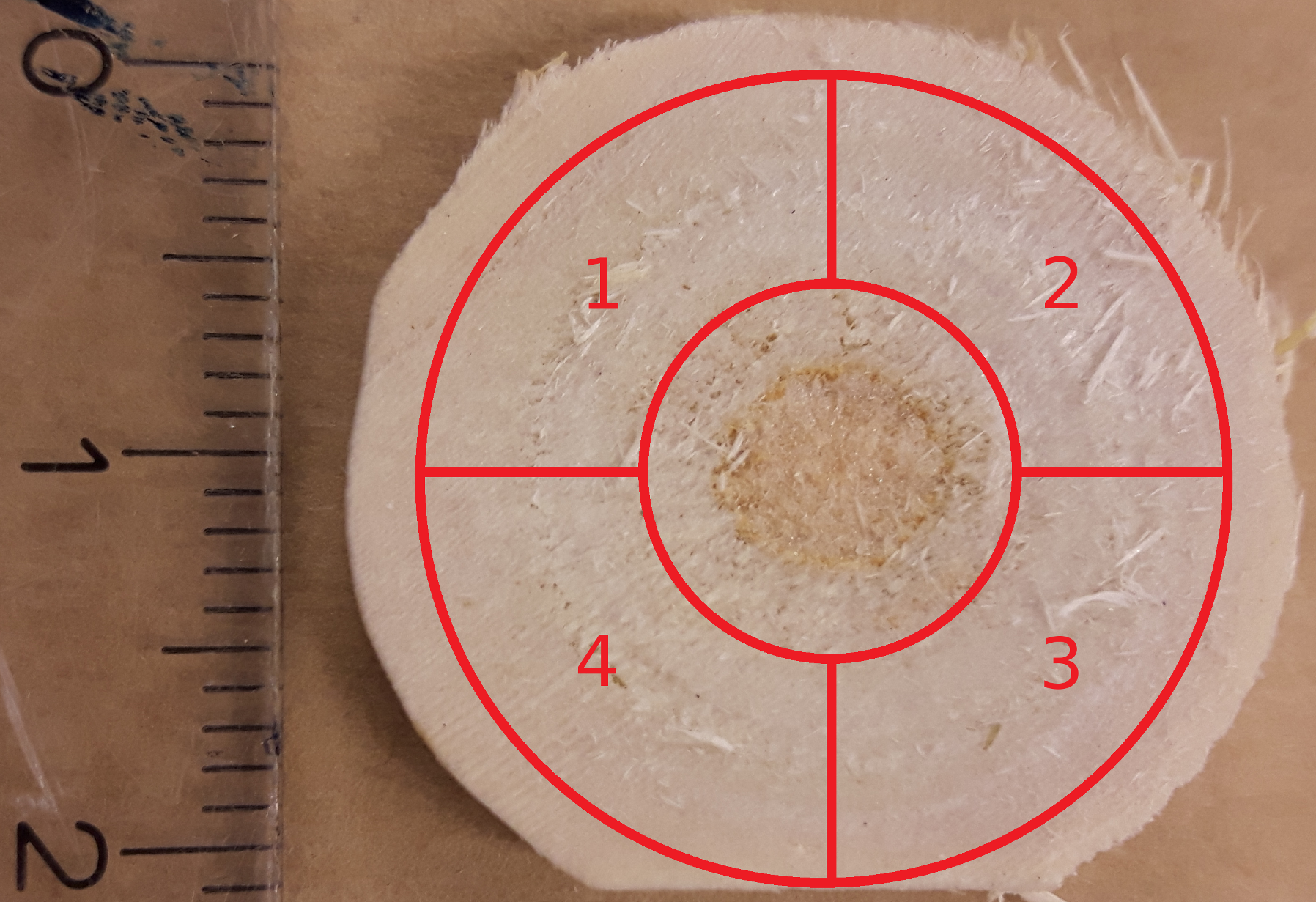}
\caption{\label{fig1}Samples for imaging were prepared according to annual rings of the willow stem wood. One of the segments (1) was dried at 60 $^\circ$C and three of them were in addition pyrolysed at (2) 308, (3) 384 or (4) 489 $^\circ$C.}
 \end{figure}

In a previous study \cite{ref19} we found that pore characteristics of ostensible same biochar may vary remarkably due to the heterogeneity of the raw materials used in biochar production. In order to minimize variation in the pore space properties originating from raw material heterogenity, following sample preparation approach was used. Fresh (moisture content 54.4$\pm$0.3 \%) willow (Salix schwerinii ‘Amgunskaja') was used as raw material. The bimodal pore size distribution of willow makes it an attractive raw material for a study considering changes is pore structure. The stem wood was first peeled and then cut into thin slices with thickness ca. 5 mm and diameter ca. 2 cm. Slices were then dried at T = 60 $^\circ$C for 48 h.  One of the thin slices was selected for imaging (Fig. \ref{fig1}). The core of the samples was removed by drilling a hole (diameter 8 mm) and the part remaining was divided in 4 segments. The outermost part of each segment was also removed so that the actual samples were approximately 5 mm wide.

\subsection{Pyrolysis}

One of the segments was only dried and three segments were in addition pyrolysed in varying temperatures. Pyrolysis was carried out with a bench-scale batch-type pyrolysis device with maximum sample size up to 500 g. The device has external electric heating system programmed to elevate temperature at speed of approximately 2 $^\circ$C min$^{-1}$. The actual temperature was measured inside the pyrolysis vessel. The device is equipped with water jacket cooler and gasbag for gas collection. Before processing, the air tightness of the device was checked and the system was thoroughly flushed with excess of N$_2$ to secure oxygen-free conditions. The holding time at peak temperature was 90 min to guarantee complete charring. The peak temperatures measured inside the pyrolysis chamber were 308, 384 and 489 $^\circ$C. Yields for biochar and liquid fractions in different process temperatures were weighted and gas volume at room temperature was measured using a gas meter (Ritter). The samples that were used in the imaging were placed in crucibles and buried within the bulk sample (300 g). Purpose of this setup was to ensure that the sample segments could be traced after pyrolysis, and that comparisons between the segments are meaningful. After pyrolysis the samples for x-ray tomography were taken from the segments. Sample size in x-ray tomography was ca. 1 mm. Imaged samples contain reasonable number of pores and they thus represent the pore structure of each segment. The experimental setup was designed to compare samples taken from a single slice of willow and thus the purpose of the samples is not to represent willow in general. More representative study containing for example several raw materials would require significantly greater number of samples, which is not currently realistic in studies using x-ray tomography (cf. e.g. recent work by Berhanu et al. \cite{ref20}). Also the samples for HIM imaging were taken from the same segments.

The rationale behind the selected process conditions is the following. The non-pyrolysed sample serves as a control and provides information of the original cellular structure of the raw material. According to Yang et al. \cite{ref21}, in pyrolysis hemicellulose degrades mainly in temperature range 220-315 $^\circ$C, cellulose in range 315-400 $^\circ$C, and lignin in wider range 160-900 $^\circ$C. Thus the lowest pyrolysis temperature demonstrates situation where hemicellulose is degraded to pyrolysis products. At the second highest temperature the major part of cellulose is deformed, and in the highest temperature, also considerable fraction of lignin has been exposed to thermochemical conversation (see \cite{ref21} for the pyrolysis curves). 

\subsection{Sample characteristics}

Before laboratory analysis, all samples (dried willow and biochars) were ground and sieved through a 2 mm sieve. Samples were analysed for loss of ignition at 550 $^\circ$C (ash content, SFS-EN 13039). The total concentrations of C, H, S and N were determined with an elemental analyzer (Flash EA1112, Thermo Finnigan, Italy), and the O content was calculated by subtracting C, H, N, S, and ash contents from the total. The BET surface area of the biochar was measured with N$_2$ sorption analysis at 77 K with a surface analyzer (TriStar II 3020, Micromeritics Instrument Corporation, USA) after degassing at 300$^\circ$C.

\subsection{X-ray microtomography and image analysis}

The 3D structural characterization of the biochars was conducted with X-ray microtomography. Zeiss MicroXCT-400 device (Pleasanton, CA, USA) was used to acquire 1600 projection images evenly distributed in full 360 degrees with exposure time of 4 seconds. The selected magnification resulted in pixel size of 1.13 $\mu$m. Source voltage was adjusted to 40 kV and source current to 250 $\mu$A. Filters were not used. 3D reconstruction was performed with device manufacturer’s XMReconstructor software.

Obtained grey-scale images were filtered and segmented as described in a previous paper \cite{ref19}. As in that paper, a fully automatic segmentation algorithm based on a modified version of Otsu's method \cite{ref22} was used to avoid operator-dependent bias in the results. This method is based on the work of Hapca et al., who added a pre-classification step to the standard Otsu method \cite{ref23}. In our variant, the average of the threshold values given by the standard Otsu method and the modified method by Hapca et al. is used, which corresponded very well with visually selected threshold value.

Images were analysed for porosity, specific surface area, pore size distribution, and structural anisotropy. A detailed account of the methods used for these analyses is given in \cite{ref19}. Here we only briefly mention that specific surface area was determined using Minkowski functionals as described by Vogel et al. \cite{ref24} and pore size distribution by method based on mathematical morphology and especially morphological opening (see \cite{ref25}). Structural anisotropy was quantified as the degree of anisotropy $D_A$ which was determined using the grey-scale gradient structure tensor \cite{ref26}. $D_A$ values close to unity describe isotropic pore structure and higher values indicate stronger structural anisotropy. In \cite{ref19}, $D_A$ was determined for several biochars derived from different feedstock and biochars with an anisotropic pore structure resulting from vascular cell structure had $D_A > 20$.

In addition to the above-mentioned analyses, three-dimensional shapes of individual pores were analysed with the method described by Claes et al. \cite{ref27}. Here all connected pore voxels were defined as a single pore and the moment of inertia tensor was computed for each set of pore voxels. The principal moments of inertia were then determined by diagonalizing the moment of inertia tensor. The pore shape can be quantified by calculating the three principal dimensions of an approximate ellipsoid using the principal moments of inertia determined for the pore. These principal dimensions $L$, $I$ and $S$ are used to characterize the pore shape, $L$ being the longest dimension of the pore, $I$ the longest dimension perpendicular to $L$, and $S$ the dimension perpendicular to $L$ and $I$. For tubular pores like the vascular pores in willow, the interesting quantity is the ratio $S/I$ which quantifies the cross-sectional shape of the pores. For circular pore cross section, this ratio is close to unity while for pores with flat cross section $S/I$ approaches zero.

\subsection{Helium ion microscopy}

HIM imaging was performed with a Carl Zeiss Orion NanoFab device. The helium ion current was 0.2-0.3 pA and the beam energy was 30 keV. Imaging of the 60 $^\circ$C and 308 $^\circ$C samples required charge neutralization with an electron flood gun between single line scans. Dwell time with these samples was 0.5 $\mu$s and a line averaging of 128 times was used. Use of the flood gun was necessary for charge neutralization as the electrical conductivity of samples were so low that significant charging effects were observed in the imaging. Samples pyrolysed at 384 $^\circ$C and 489 $^\circ$C did not require the use of the flood gun due to their higher conductivity. The dwell time with these samples was 10 $\mu$s and the line averaging was 4 times. Pressure in the measurement chamber during the imaging was 1-2$\cdot$10$^{-7}$ Torr.

\section{Results and discussion}

Basic properties of the dried willow and biochars are shown in Table \ref{table1}. The clear decrease in O/C and H/C ratios with increasing pyrolysis temperature indicates changes in the chemical composition and structural arrangement of the initial organic matter. The biochars produced at 384 and 489 $^\circ$C had O/C ratio less than 0.2 which indicate high degree of carbonization. The high H/C ratio of the biochar produced at 308 $^\circ$C indicates existence of non-condensed aromatic structures (e.g. lignin) \cite{ref28}. When applied in soil, decrease in O/C and H/C ratios indicates higher stability against microbiological degradation thus improving carbon sequestration potential of biochar.

\begin{table*}
\caption{
Elemental composition, ash content , H/C and O/C atomic ratios, and BET surface area of dried and pyrolysed willow stem wood samples.
}\label{table1}
\begin{tabular}{ l c c c c c }
  \hline\noalign{\smallskip}
   & Dried & & & Pyrolysed &  \\
   & 60 $^\circ$C & & 308 $^\circ$C & 384 $^\circ$C & 489 $^\circ$C \\
  \noalign{\smallskip}\hline\noalign{\smallskip}

  C [wt-\%] & 44.6 & & 67.9 & 75.9 & 83.7 \\
  H [wt-\%] & 5.78 & & 4.72 & 4.00 & 3.18 \\
  N [wt-\%] & 0.15 & & 0.27 & 0.31 & 0.37 \\
  S [wt-\%] & 0.02 & & 0.01 & 0.01 & 0.01 \\
  O [wt-\%] & 48.9 & & 25.9 & 18.2 & 8.3 \\
  H/C ratio & 1.54 & & 0.83 & 0.63 & 0.45 \\
  O/C ratio & 0.82 & & 0.29 & 0.18 & 0.07 \\
  Ash [wt-\%] & 0.51 & & 1.19 & 1.60 & 4.44 \\
  BET [m$^2$g$^{-1}$] & n.d. & & 3.96 & 7.47 & 6.82 \\
  \noalign{\smallskip}\hline
\end{tabular}
\end{table*}

The image analysis data (Table 2) indicate that pyrolysis increased micrometre-scale porosity (9-16 \%) and specific surface area (7-13 \%) of biochars compared to those of dried willow stem wood. Note that the starting point in our study is dried willow which has already shrunken as compared to fresh willow. Change of pyrolysis peak temperature in the range 308-489 $^\circ$C had only minor influence on the porosity and specific surface area. All $D_A$ values are at high level irrespectively of the treatment (reference data for typical $D_A$ values determined for various biochar structures can be found from \cite{ref19}).

\begin{figure*}
\includegraphics[width=0.75\textwidth]{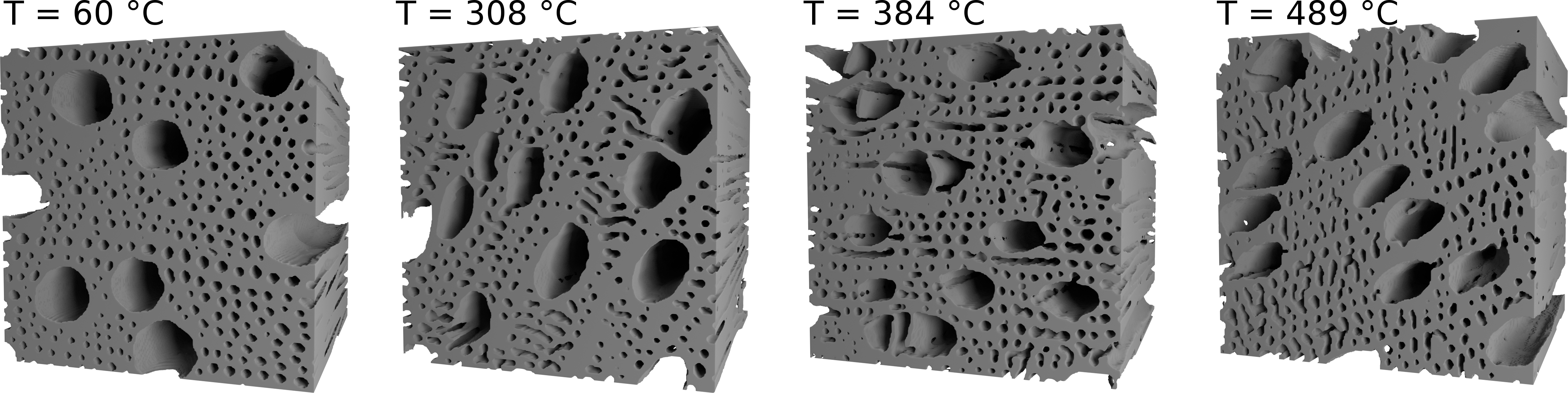}
\caption{\label{fig2}3D visualizations of the x-ray tomographic images. Size of each visualized subsample is 280 $\mu$m$\times$280 $\mu$m$\times$280 $\mu$m.
}
\end{figure*}

These results, together with visual inspection of the pore structure (Fig. \ref{fig2}), suggest that the inherent vascular cell structure of willow determines the micrometre-scale pore structure of the biochar. Pyrolysis can modify inherent pore structure of raw material to some extent. These changes can be obtained already in the temperature typical for torrefaction, however, the pore structure remained relatively stable over the pyrolysis temperature regime studied here. High degree of anisotropy (high $D_A$ value), both in wood and biochar samples, indicates straight pipe-like form of the pore system with poor connectivity of the pore system in lateral directions. On the other hand, pipe-like structure ensures good connectivity of the internal porosity to the surfaces of biochar particles.

\begin{figure*}
\includegraphics[width=0.75\textwidth]{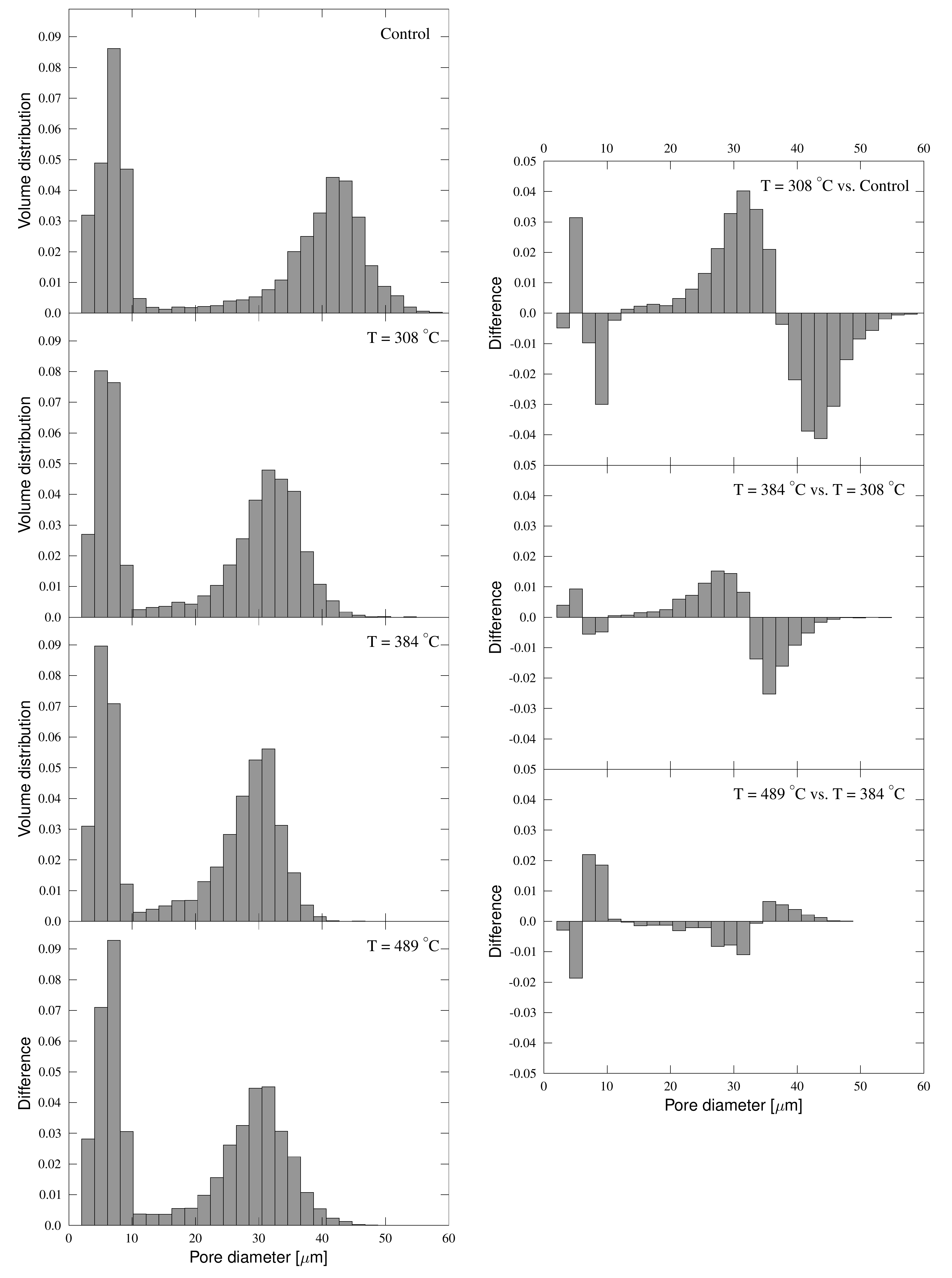}
\caption{\label{fig3}Pore size distributions on left and differences of two adjacent distributions on right. 
}
\end{figure*}

Closer inspection of the pore size distributions (Fig. \ref{fig3}) shows that changes observed in the total porosity and SSA arise from the shrinkage of biochars. This means that pore size distribution shifts toward smaller pore sizes when sample undergoes pyrolysis process. As can be seen in the difference plots (panels on the right-hand side in Fig. \ref{fig3}), most considerable change in the pore size distribution takes place between dried sample and lowest pyrolysis temperature (308 $^\circ$C). Pore size distribution was shifted slightly towards smaller pore sizes between 308 and 384 $^\circ$C. Between the two highest pyrolysis temperatures (384 and 489 $^\circ$C) a tiny shift towards larger pore sizes was observed. The differences between different pyrolysis temperatures are most probably consequences of the accuracy of imaging and pore-size distribution analysis and do not reflect true changes in the pore structure.  

The cross-sectional shape of the pores was found to be different for dried willow and biochars (Table \ref{table2}. Lower $S/I$ value for biochars indicate that shrinkage of the raw material in pyrolysis is anisotropic whereby the cross-sectional shape of the pores is more elongated in biochars than in non-pyrolysed willow. Visual inspection of pore shapes supports this finding (Fig. \ref{fig2}). Shape analysis was done separately for large and small pores and the results for both pore types are consistent. It seems that for small pores, the difference in pore shape may partly result from connections formed between adjacent pores (Fig. \ref{fig2}). However, as the imaging resolution is fairly close to the thickness of pore walls, this conclusion is uncertain.

\begin{table*}
\caption{
Results of 3D image analysis. Porosity, specific surface area (SSA), degree of anisotropy ($D_A$) and shape ratio $S/I$ determined for dried willow and biochars pyrolysed at different temperatures. For $S/I$ ratio, mean value and standard deviation were calculated separately for large and small pores.
}\label{table2}
\begin{tabular}{ l c c c c c }
  \hline\noalign{\smallskip}
   Sample & Porosity [-] & SSA [mm$^2$mm$^{-3}$] & $D_A$ [-] & $S/I$ (large pores) [-] & $S/I$ (small pores) [-]  \\
  \noalign{\smallskip}\hline\noalign{\smallskip}
  Dried willow              & 0.32 & 98  & 88 & 0.77$\pm$0.12 & 0.77$\pm$0.14 \\
  Pyrolysis at 308 $^\circ$C & 0.36 & 105 & 48 & 0.61$\pm$0.20 & 0.63$\pm$0.0.21 \\
  Pyrolysis at 384 $^\circ$C & 0.35 & 110 & 82 & 0.45$\pm$0.15 & 0.61$\pm$0.24 \\
  Pyrolysis at 489 $^\circ$C & 0.37 & 111 & 30 & 0.60$\pm$0.14 & 0.62$\pm$0.22 \\
  \noalign{\smallskip}\hline
\end{tabular}
\end{table*}

\begin{figure*}
\includegraphics[width=0.75\textwidth]{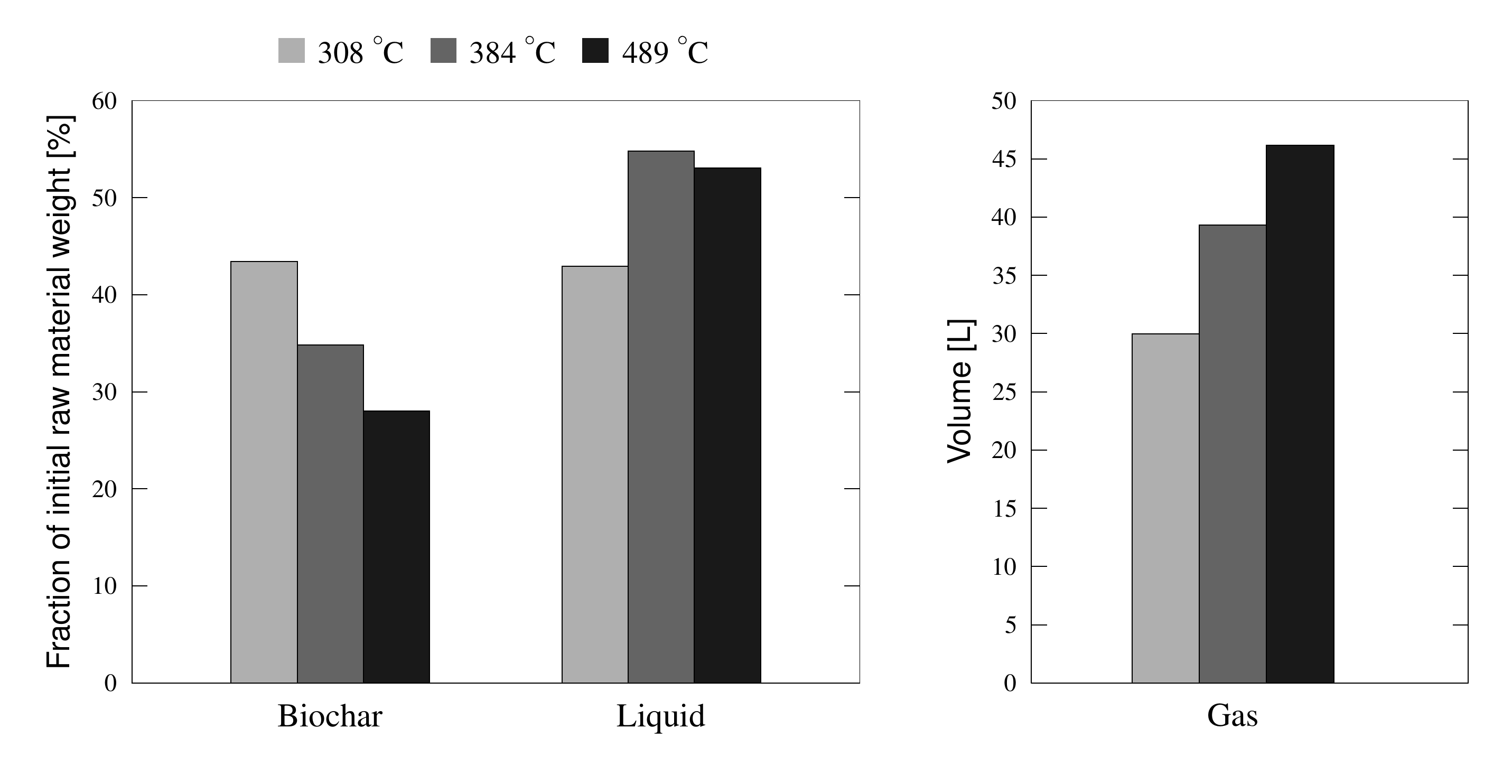}
\caption{\label{fig4}Yields for biochar, liquid and gas fractions in different process temperatures. Left panel shows the fraction of initial raw material weight converted to biochar and liquid and right panel the volume of produced gas.
}
\end{figure*}

The yield data presented in Fig. \ref{fig4} explain why there are major differences between the dried sample and the lowest pyrolysis temperature. In this first step as much as 57 \% of initial mass of the dried raw material was converted into gases and liquids. The additional mass loss from the lowest pyrolysis temperature to the highest one is only 15 \%. However, it is notable that at the highest pyrolysis temperature roughly three-fourths (72 \%) of total raw material weight is lost (i.e. converted into gases and liquids), but still the micrometre-scale pore structure of biochar retains relatively well its initial geometry and size.

The next question is that how does weight loss of 72 \% affect the biochar quality? BET results (Table \ref{table1}) indicate very low surface area for all studied biochars. In line with general assumption, there is a slight increase in measured surface area with increasing peak pyrolysis temperature. The minimal increase of surface area in the temperature regime considered here is expected as previous studies have shown that greater increase happens at higher temperature (see e.g. \cite{ref12,ref29}). BET surface areas show a slight increase in surface from lowest to highest process temperature, which is in principle similar than that seen in image analysis results (Table \ref{table2}). However, it must be noted that these two techniques probe different pore size regime and thus the results are not comparable.

\begin{figure*}
\includegraphics[width=0.75\textwidth]{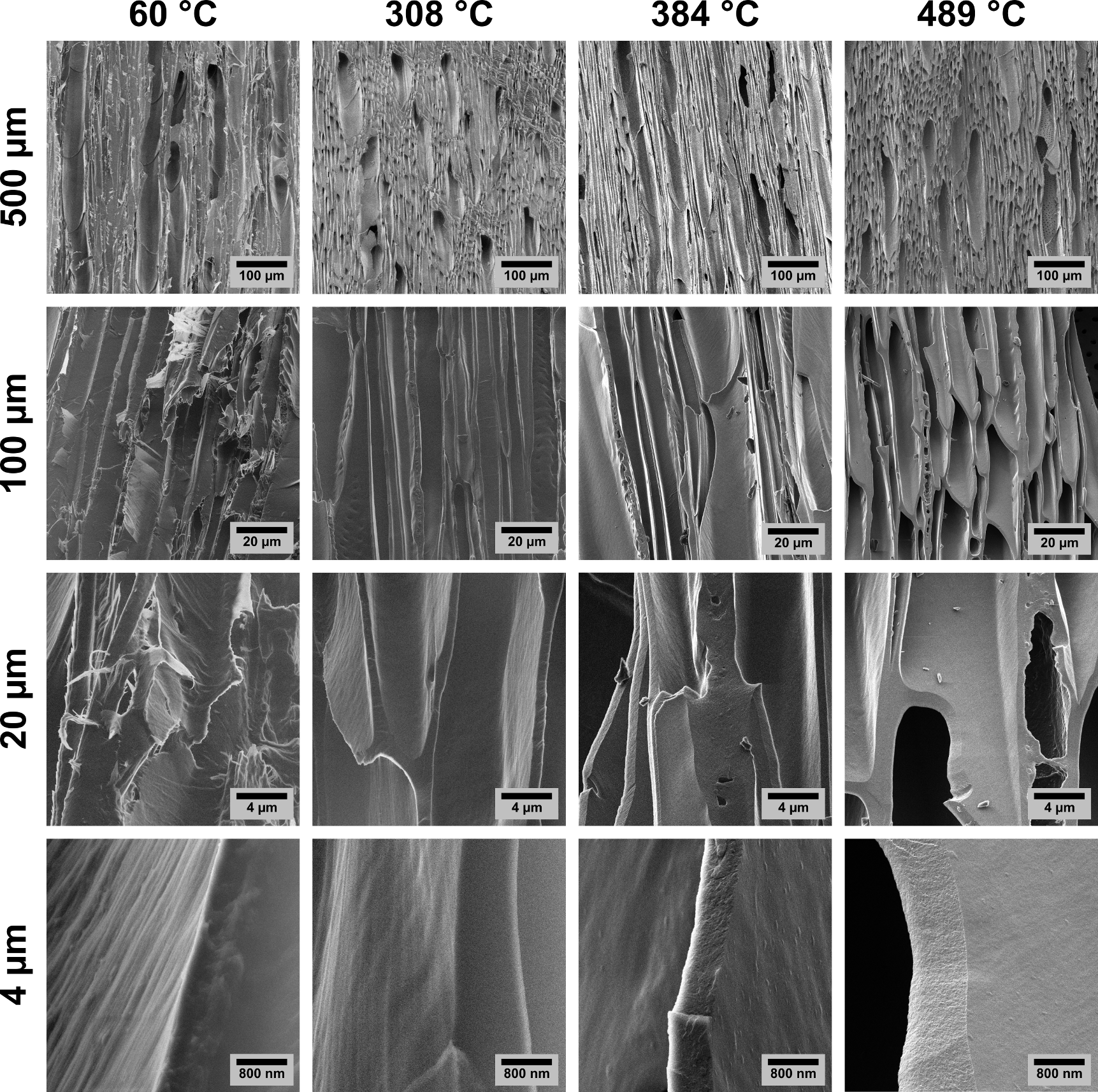}
\caption{\label{fig5}Helium ion microscopy images of studied samples in four different image sizes. From left to right, dried willow and biochars pyrolysed at increasing temperature. Size of the images increase from 4$\times$4 $\mu$m$^2$ to 500$\times$500 $\mu$m$^2$ from bottom to the top. 
}
\end{figure*}

Gas adsorption results are supported by observations made with HIM. HIM images of the four sample types are shown in Fig. \ref{fig5}. In spite of the very high resolution of HIM images, significant differences in the surfaces of cellular pores cannot be observed indicating that the mass loss in pyrolysis generates structures mainly at subnanometre length scales. This is in agreement with the theory presented in \cite{ref7}, which suggests that additional porosity would result from structural defects in carbon. Comparison of the HIM images and the x-ray tomography visualizations show similar vascular pore structures of two different size groups, namely one of about 5 µm in diameter and another 30-50 $\mu$m in diameter.  Highest resolution HIM images also show slightly more sub-micron roughness in the pore walls of the 60 $^\circ$C dried sample compared to the smoother inner surfaces of the pyrolysed samples.

High-resolution HIM images show that x-ray tomography visualization exaggerates the wall thickness between the smaller diameter pores. The wall thickness is close to the resolution of x-ray tomography, whereby segmentation of walls becomes inaccurate. While resolution of x-ray tomography was not able to resolve pores in submicrometre size range which partially contributes to the porosity able to store plant available water, HIM images ensure that the studied biochar did not contain such porosity.

The pyrolysis temperature mainly affects chemical properties instead of physical ones (Table 1). Thus, when biochar use as soil amendment is considered, selection of process temperature should be judged on the basis of chemical measures. For example, acidic functional groups on biochar surface contribute to their wettability and water retention capacity [8]. The relationship between capillary pressure across the water meniscus ($\Delta p$), pore diameter ($d$) and contact angle ($\theta$) is described by the Young-Laplace equation,
\begin{equation}\label{eq1}
\Delta p = \frac{4\gamma \cos \theta}{d},
\end{equation}
where $\gamma$ is the surface tension of water \cite{ref4}. Thus increasing pyrolysis temperature can alter the water retention properties either by changing the pore size distribution (i.e., pore diameter in Eq. (\ref{eq1})) or surface wettability via surface chemistry (contact angle). According to Young-Laplace equation, reduced wettability (increased $\theta$) decreases the pore volume which is able to store water at a given soil-water potential.  

Our results do not necessarily indicate that pyrolysis temperature would not have any influence on the water retention properties of biochars and biochar-amended soils. However, if such effects exist, the explanation would be in the modified surface chemistry rather than in the physical properties related to modified pore structure. Hydrophobicity of biochar decreases with increasing pyrolysis temperature \cite{ref30} and biochar can be either hydrophobic or hydrophilic.  

It is also important to note that pyrolysis temperature has an effect on the cation exchange capacity of biochar \cite{ref31}, and the change in this property may also depend on the process temperature \cite{ref32,ref33}. As cation exchange capacity affects soil fertility, it should be also accounted for when optimizing process parameters for production of biochars for soil amendment purposes.

Concerning soil amendment use of biochar, our study is limited to so-called direct effects of biochar, where biochar pores directly act as storage space for water \cite{ref34}. Biochar may in addition also have indirect effects, as biochar application may affect the water retention properties of soil by improving soil structure, aggregation and aggregate stability \cite{ref34,ref35,ref36,ref37,ref38}. Also the surface chemistry of biochar may have direct or indirect effects. Biochar contact angle affects directly the water retention in biochar particles (as described by Eq. (\ref{eq1})). In addition, depending on the chemical characteristics, biochar amendment may result to formation of water repellent coatings on soil aggregates [39] or remove hydrophobic compounds from the soil \cite{ref40} which affects the water repellency of the biochar amended soil \cite{ref41}.

In this work we considered only relatively low process temperatures due to device-specific limitations. Considering future work, similar approach should be applied to temperature range covering also higher pyrolysis temperatures as well as to raw materials with different chemical properties. The effects of heating rate were not considered here, but it has been found to influence the porosity development at nanometre size range but only minor effect on the density of surface functional groups \cite{ref42}.

\section{Conclusions}

Our results indicate that at pyrolysis temperatures below 500 $^\circ$C and with low heating rate (2 $^\circ$C min$^{-1}$), the initial pore structure of raw material (in this work willow) determines the micrometre-scale porosity of biochar. Considering the pore size regime relevant for plant water uptake, pyrolysis temperature cannot be used to optimize the micrometre-scale pore system of biochar. Also, the nanoscale porosity of biochar remained at low and virtually constant level when subjected to process temperatures less than 500 $^\circ$C, which is typical for torrefaction and slow pyrolysis.

Considering the development of engineered biochars for agronomic applications with specific target to improve soil water holding capacity, following conclusions can be drawn: (1) If aim is to increase the soil water holding capacity at specific matric potential regime directly via biochar pores, selection of appropriate raw material with desired pore size distribution is essential; (2) In such application it should, however, be verified how much torrefaction or low temperature pyrolysis modifies pore size distribution via shrinkage of raw material; (3) If porosity remains at satisfactory range, process conditions can be optimized for desired chemical properties of biochar (e.g. surface functional groups, hydrophobicity, carbon sequestration potential), quality of derived pyrolysis liquids, and/or energy balance of the entire process. 

\begin{acknowledgments}
This project has received funding from the European Union’s Horizon 2020 research and innovation programme under grant agreement No 637020 - MOBILE FLIP.
\end{acknowledgments}

\end{document}